\documentclass[preprint,12pt]{elsarticle}

\usepackage{amssymb}
\usepackage{epsfig}
\usepackage{graphics}
\usepackage{graphicx}
\usepackage{subfigure}
\usepackage{exscale}
\usepackage{latexsym}
\usepackage{makeidx}
\usepackage[section]{placeins}
\usepackage{xtab}
\usepackage{epsf}
\usepackage{bbm}
\usepackage{color}

\journal{Nuclear Instruments and Methods in Physics Research Section A}

\begin{document}
\begin{frontmatter}

\title{Response of diamond detector sandwich to 14 MeV neutrons}

\author{M.~Osipenko$^a$, 
M.~Ripani$^a$, G.~Ricco$^a$, B.~Caiffi$^a$,
F.~Pompili$^b$, M.~Pillon$^b$,
G.~Verona-Rinati$^c$,
R.~Cardarelli$^d$
}

\address{
$^a$ \it\small INFN, sezione di Genova, 16146 Genova, Italy, \\
$^b$ \it\small ENEA, Frascati, 00044 Italy. \\
$^c$ \it\small Universit\`a di Tor Vergata, Rome, 00133 Italy. \\
$^d$ \it\small INFN sezione di Roma II, 00133 Italy. \\
}

\begin{abstract}
In this paper we present the measurement of the response of 50 $\mu$m thin diamond detectors
to 14 MeV neutrons. Such neutrons are produced in fusion reactors and are of particular
interest for ITER neutron diagnostics.
Among semiconductor detectors diamond has properties most appropriate for harsh radiation and
temperature conditions of a fusion reactor. However, 300-500 $\mu$m thick diamond detectors suffer
significant radiation damage already at neutron fluences of the order of $10^{14}$ n/cm$^2$.
It is expected that a 50 $\mu$m thick diamond will withstand a fluence of $>10^{16}$ n/cm$^2$.
We tested two 50 $\mu$m thick single crystal CVD diamonds,
stacked to form a ``sandwich'' detector for coincidence measurements.
The detector measured the conversion of 14 MeV neutrons, impinging on one diamond, into $\alpha$ particles
which were detected in the second diamond in coincidence with nuclear recoil.
For $^{12}C(n,\alpha)^{9}Be$ reaction the total energy deposited in the detector
gives access to the initial neutron energy value.
The measured 14 MeV neutron detection sensitivity through this reaction
by a detector of effective area 3$\times$3 mm$^2$ was $5\times 10^{-7}$ counts cm$^2$/n.
This value is in good agreement with Geant4 simulations.
The intrinsic energy resolution of the detector was found to be 240 keV FWHM
which adds only 10\% to ITER's 14 MeV neutron energy spread.
\end{abstract}


\end{frontmatter}

\section{Introduction}\label{sec:intro}
Diamonds were proposed as one of detectors for the Radial Neutron Camera (RNC) diagnostics of ITER~\cite{rnc_design}.
The preliminary design~\cite{mosaic_cvd} foresees the construction of a matrix of single diamond crystals
with nominal thickness of 500 $\mu$m. The choice of the detector is related to the harsh
environmental conditions inside ITER's port. The most relevant include high neutron flux,
of the order of $10^9$ n/cm$^2$s, high operational temperature 100 C$^\circ$ and large $\gamma$ and X-ray background.
Diamond, indeed, features an order of magnitude higher radiation hardness with respect to Silicon.
It is less sensitive to $\gamma$s and X-rays in comparison to other semiconductors.
And diamonds can operate at high temperatures without dramatic build-up of intrinsic noise.
All these properties make the diamond best suitable fast neutron detector for ITER diagnostics
among semiconductor sensors.

However, even diamond detectors suffer from radiation damage by fast neutrons, which cannot be totally neglected.
In particular, RNC of ITER is supposed to operate for many years without any possibility
of detector replacement. According to operation scenario the diamond detector performance
can deteriorate before the end of operations. Many studies of diamond radiation hardness
have been conducted so far. However, experimental tests with 14 MeV neutrons are scarce and incomplete.
Most detailed studies were performed with proton beams by RD42 Collaboration~\cite{kagan_rd42}.
According to the measured parametrization of charge collection distance (CCD) as a function
of proton fluence the 500 $\mu$m thick diamond loses 10 \% of the signal already after $5\times 10^{14}$ p/cm$^2$.
The same parametrization suggests that a 50 $\mu$m thick diamond will lose 10 \% of the signal
after $3\times 10^{16}$ p/cm$^2$, while a 20 $\mu$m thick diamond will withstand $8\times 10^{16}$ p/cm$^2$.
These numbers cannot be directly applied to 14 MeV neutrons because the damage
produced by high energy protons is different. Nevertheless, assuming that the damage
is proportional to the non-ionizing energy loss this difference is less than a factor of two~\cite{cvd_rad_hard}.

\begin{figure}[!ht]
\begin{center}
\includegraphics[bb=2.5cm 1.5cm 20cm 27.5cm, scale=0.35, angle=270]{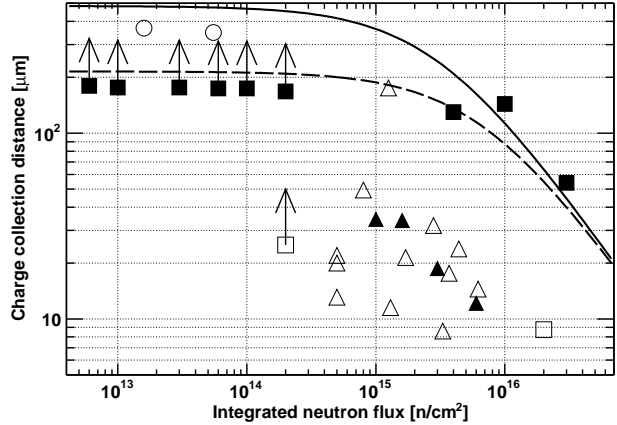}
\caption{\label{fig:ccd_flux}Charge collection distance in diamond detectors as a function of
irradiating neutron fluence.
The data are from Refs.~\cite{allers,oh,bergonzo,bruzzi,alekseev,tanaka,muller,pillon,almaviva,lohstroh}.
Triangles represent the data taken on polycrystalline diamonds, where the empty triangles
are taken with neutrons of energy above 3 MeV, while full triangles with fission neutrons.
Squares show the data obtained on single crystal CVD or natural diamonds, where empty squares
indicate the diode-like detectors using p-type diamond as one contact.
Empty circles show the data taken on HPHT single crystal diamonds.
Solid and dashed lines give the CCD parametrization~\cite{kagan_rd42} for high energy proton irradiation
of single crystal and polycrystalline diamonds, respectively.
}
\end{center}
\end{figure}

Neutron irradiation studies were performed
in Refs.~\cite{allers,oh,bergonzo,bruzzi,alekseev,tanaka,muller,pillon,almaviva,lohstroh}
as summarized in Fig.~\ref{fig:ccd_flux}.

In Ref.~\cite{allers} a 35 $\mu$m polycrystalline diamond detector was irradiated
with $6\times 10^{15}$ n/cm$^2$ fast neutrons at VIPER (UK) fission reactor.
The irradiation resulted in the reduction of charge collection by a factor of 3.

In Ref.~\cite{oh} 50, 90 and 545 $\mu$m polycrystalline diamond detectors were irradiated
with $6.5\times 10^{15}$ n/cm$^2$ at PTB (Germany) 20 MeV deuteron induced fast neutron source (thick beryllium target).
The neutron mean energy was 5 MeV.
The irradiation resulted in the reduction of charge collection to 60\%, 30\% and 20\%, respectively.

In Ref.~\cite{bergonzo} a 25 $\mu$m polycrystalline diamond detector was irradiated
with $3\times 10^{15}$ n/cm$^2$ at a CEA fast fission reactor.
The absorbed dose resulted in 25\% decrease of the collected charge.
CCE before irradiation was not measured and assumed to be 100 \%.

In Ref.~\cite{bruzzi} a 600 $\mu$m polycrystalline diamond detector was irradiated
with $5\times 10^{14}$ n/cm$^2$ at ATOMKI (Hungary) 16 MeV proton induced fast neutron source (thick beryllium target).
The neutron mean energy was 4 MeV.
The irradiation resulted in the reduction of charge collection by a factor of 8.
CCE before irradiation was not measured and assumed to be 100 \%.

In Ref.~\cite{alekseev} a 100 $\mu$m natural IIa diamond detector was irradiated
with $3\times 10^{16}$ n/cm$^2$ at IR-8 (Russia) fission reactor.
The charge collection curve as a function of accumulated flux shows stable response until $10^{14}$ n/cm$^2$,
with a drop to 30\% after $3\times 10^{16}$ n/cm$^2$.

In Ref.~\cite{tanaka} a 190 $\mu$m HPHT single crystal diamond detector was irradiated
with $5.5\times 10^{13}$ n/cm$^2$ 14 MeV neutrons from FNS (Japan) DT generator.
The irradiation resulted in the reduction of charge collection on 10\%.
CCE before irradiation was not measured and assumed to be 100 \%.

In Ref.~\cite{muller} a 350 $\mu$m polycrystalline mechanical grade diamond detector was irradiated
with $1.25\times 10^{15}$ n/cm$^2$ at Louvain la Neuve neutron source with a mean neutron energy of 20 MeV.
A factor of two charge loss was observed after this dose.
CCE before irradiation was not measured and assumed to be 100 \%.

In Ref.~\cite{pillon} a 25 $\mu$m single crystal diamond was irradiated
with $2\times 10^{14}$ n/cm$^2$ 14 MeV neutrons from FNG (Frascati, Italy) DT generator.
No significant ($>1$\%) change of charge collection was seen.

In Ref.~\cite{almaviva} a 25 $\mu$m single crystal diamond was irradiated
with $2\times 10^{16}$ n/cm$^2$ fission neutrons from TRIGA reactor (Casaccia, Italy).
35\% charge collection was observed at the end of irradiation.

In Ref.~\cite{lohstroh} a 300 $\mu$m single crystal diamond detector was irradiated
with $1\times 10^{16}$ n/cm$^2$ fast neutrons at VIPER (UK) fission reactor.
The irradiation resulted in the reduction of charge collection to 18\%.

It has to be noticed that some of these measurements do not provide a complete information.
In particular, the data from Refs.~\cite{bruzzi,tanaka,muller}
do not give charge collection efficiency before irradiation,
which can be as low as 10\% for polycrystalline and HPHT crystals.
This explains why in Fig.~\ref{fig:ccd_flux} the points from Ref.~\cite{tanaka} and~\cite{muller}
lie above the single crystal values.
Moreover, the author of Ref.~\cite{muller} used a mechanical grade crystal,
whose CCE is known to be inferior to those of electronic grade crystals.
Another difference lies in the definition of CCD measured in Refs.~\cite{alekseev,tanaka,lohstroh}
with $^{241}Am$ source, whose $\alpha$ particle range in diamond is smaller than diamond thickness.
The correction on these data points, roughly a factor of 2, is included in Fig.\ref{fig:ccd_flux}.
The single crystal results are mostly limited to Ref.~\cite{alekseev} and~\cite{lohstroh},
which are found to be in good agreement and define a clear function, very similar to the proton CCD curve.
The diode-like single crystal detectors give significantly lower CCD, probably
due to faster damage of p-type layer.
Therefore, all these data do not contradict the above mentioned statements.

The existing data indicate that if RNC of ITER were to require a signal stability of the neutron detector
up to fluences of 10$^{16}$ n/cm$^2$, diamonds of small thickness have to be used.
This clearly reduces the detection efficiency, proportional to the sensor's active volume,
by a factor of 25 for a 20 $\mu$m thick diamond.
But, given the expected $^{12}C(n,\alpha)^{9}Be$ event rate of 6 kHz on a single sensor,
it still allows to meet the requirement~\cite{rnc_design} of 10\% statistical precision
in 1 ms period combining rates of a 4x4 matrix.
The reduction of diamond thickness also increases the noise due to higher sensor capacity and
larger dark current. However, the noise from readout electronics remains the dominant contribution,
in particular if the first amplifier cannot be installed near the sensors.
Some of the noise can be suppressed together with the background from elastic n-C scattering by measuring
coincidences between two stacked diamond detectors.
In this case 20 $\mu$m represents the maximum diamond depth from which an $\alpha$ could emerge
triggering the coincidence.
It also has to be considered that at ITER the intrinsic
14 MeV neutron peak FWHM is expected to be about 500 keV. Thus, defining how large the instrumental contribution
to the energy resolution would be acceptable one can estimate the needed requirements.
For example if the detector energy resolution has to be $<10$\% of the total
this gives a 230 keV FWHM limit.

In this article we describe the measurement of the response of a diamond sandwich detector
to 14 MeV neutrons produced by a DT generator.
Some details of the detector design are given in section~\ref{sec:det}.
The measurement of the detector response to 14 MeV neutrons is discussed in section~\ref{sec:fng},
while the comparison to Geant4 simulations and its implications are given in section~\ref{sec:res}.

\section{Detector}\label{sec:det}
The sandwich detector was made of two Single-Crystal Diamond detectors (SCD),
labeled as SCD282 and SCD1517, produced at the University of Rome ``Tor Vergata''.
The detailed description of the detector is given in Ref.~\cite{sdw_calib}.

Each SCD has a total area of $4\times 4$ mm$^2$ and is composed of five layers:
HPHT substrate, p-type diamond, intrinsic diamond, metallic contact~\cite{fulvio_r1,fulvio_r2} and LiF layer.
The highly doped p-type layer acts like an ohmic contact, and the metallic anode
is connected to ground.

The metallic anode on the intrinsic diamond was applied by evaporation of a 3 mm $\times$ 3 mm $\times$ 40 nm Chromium layer,
which was also used as a sticking layer for two narrow (0.4 mm $\times$ 3 mm $\times$ 80 nm) gold strips.
These strips were used to connect the anode to ground via mechanical contact with the two 50 $\mu$m thick microwires
stretched along the strips.
A layer of LiF converter was then deposited on top of the chromium layer in between the gold strips
and had dimensions of 3 mm $\times$ 2.2 mm x 50 nm.
The active volume of each SCD consists of its intrinsic diamond layer located under its $3\times3 $ mm$^2$ metallic contact.
Outside of the metallic contact area the electric field is missing precluding charge collection.
The thicknesses of the Boron doped (p-type) and intrinsic diamond layers were different for the two SCDs,
as shown in Table~\ref{tab:scd_character}.

\begin{table}[!h]
\begin{center}\label{tab:scd_character}
\caption{Characteristics of selected diamond sensors.} \vspace{2mm}
\begin{tabular}{|c|c|c|c|c|} \hline
  SCD   & P-doped  & Intrinsic & Cr      & LiF   \\ 
        & layer    &  layer    & contact & layer  \\ 
        & thick.   & thick.    & thick.  & thick. \\ 
        & [$\mu$m] & [$\mu$m]  & [nm]    & [nm]   \\ \hline
SCD282  & 26       & 54        & 40      & 50    \\ \hline
SCD1517 & 15       & 49        & 40      & 50    \\ \hline
\end{tabular}
\end{center}
\end{table}

The two SCDs are placed one on top of the other with their metallic contacts facing each other (``sandwich'' structure).
The positive signals are collected from the p-type layer, connected with a droplet of conductive glue
to the housing PCB traces transporting the signal to the pole wire of the outgoing cables.

\begin{figure}[h]
\begin{center}
\includegraphics[bb=1cm 4cm 21cm 26cm, scale=0.2]{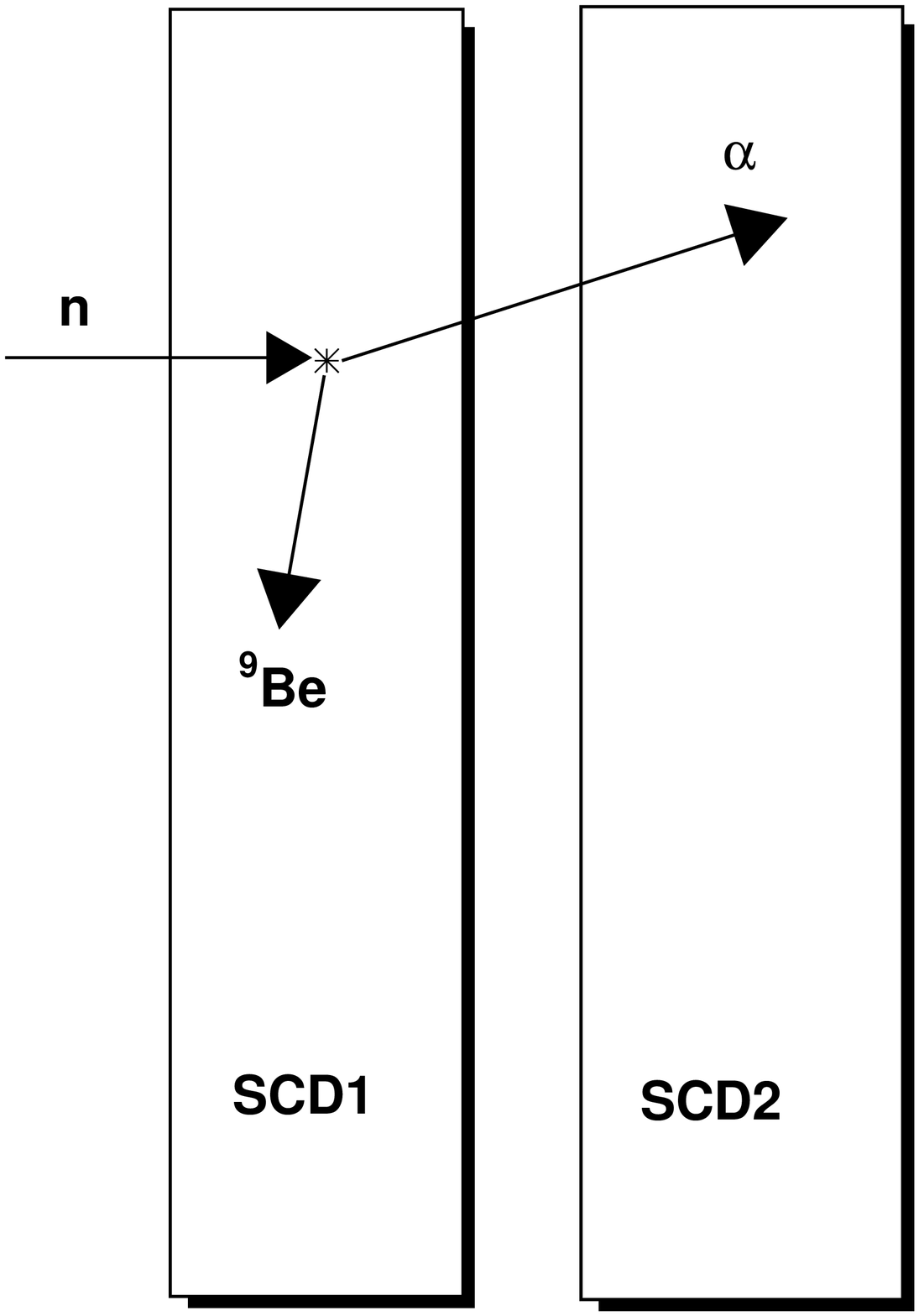}%
\includegraphics[bb=1cm 4cm 21cm 26cm, scale=0.2]{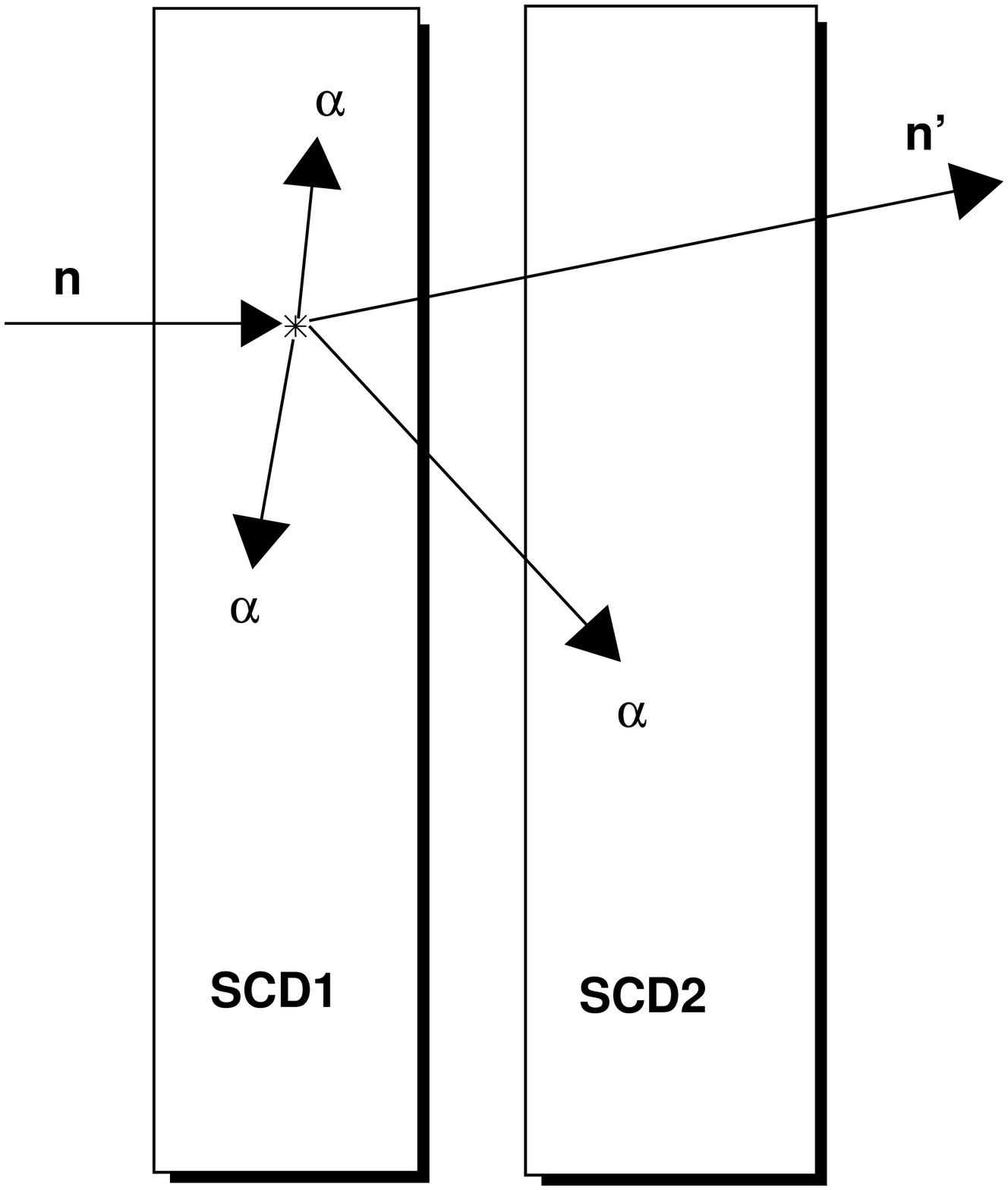}%
\includegraphics[bb=1cm 4cm 21cm 26cm, scale=0.2]{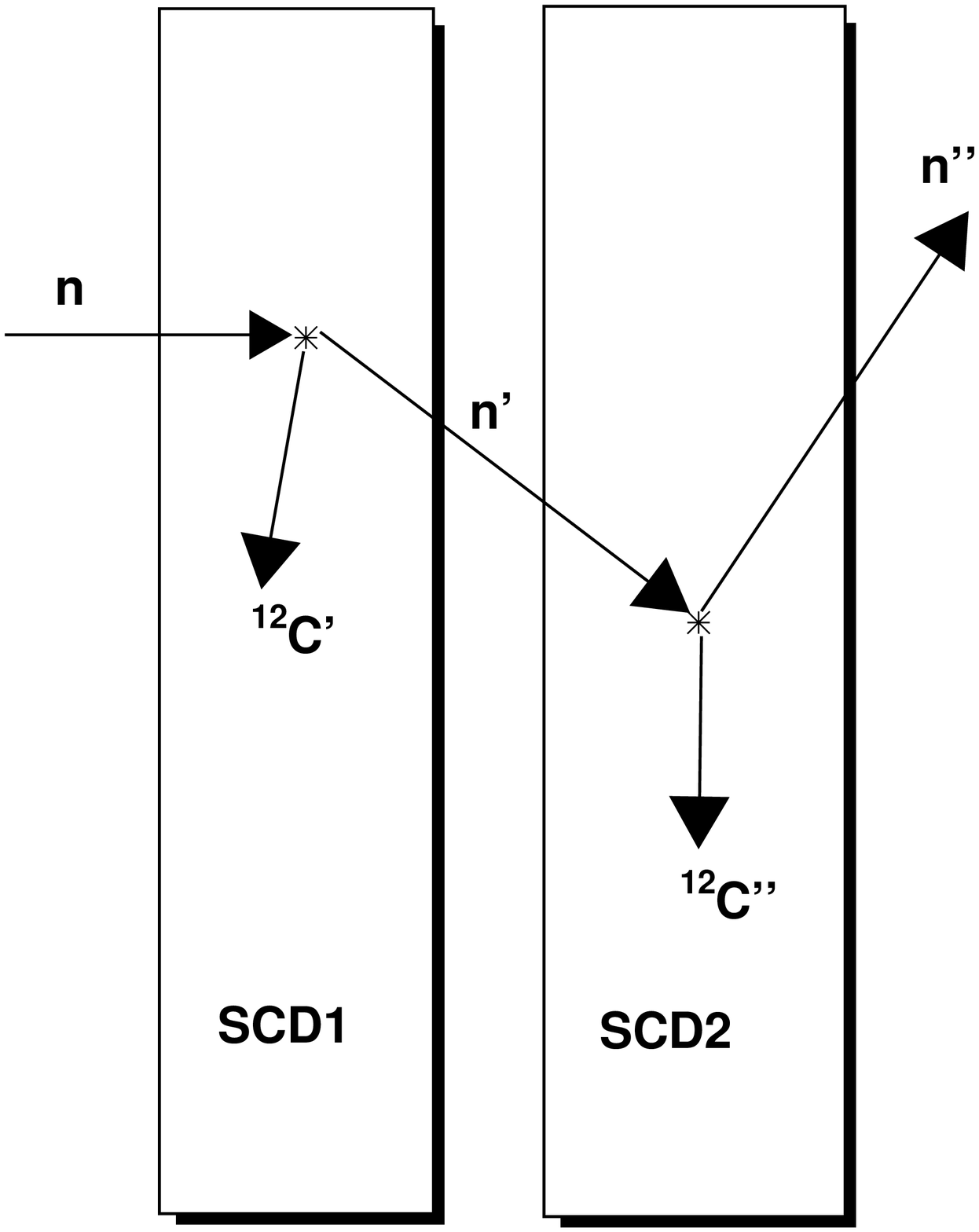}
\caption{\label{fig:sdw_reac}Interactions of 14 MeV neutrons with sandwich detector
resulting in measurable coincidences. From left to right:
$^{12}C(n,\alpha)^{9}Be$,
$^{12}C(n,n')3\alpha$,
and $^{12}C(n,n')^{12}C^*$.}
\end{center}
\end{figure}

The detection of the 14 MeV neutrons is performed in coincidence between the two SCDs
through $^{12}C(n,\alpha)^{9}Be$ reaction in the first diamond's bulk,
where the produced $\alpha$ traveled into the second diamond.
These coincidences may occur through the reactions shown in Fig.~\ref{fig:sdw_reac}.
To model correctly these reactions it is important to know
the active volume of each SCD and the material thickness between the diamonds crossed by the $\alpha$ particles.
At variance with 14 MeV neutrons, thermal neutrons are measured through their conversion in $^6$Li isotope (96\% abundance in the LiF used).
The $\alpha$ and $t$ emitted back-to-back penetrate each into one of the intrinsic diamond depletion layers
and trigger the coincidence.

For detection of the 14 MeV neutrons through $^{12}C(n,\alpha)^{9}Be$ reaction in coincidence
it is sufficient to have 20 $\mu$m thick diamond crystals. Indeed, Geant4 simulations
performed in this study indicated that the $\alpha$ particles produced in one diamond
can leave its volume and enter the second diamond, triggering the coincidence,
only if they are generated at $<20$ $\mu$m depth from the metallic contact.
The use of 50 $\mu$m crystals leads to an increase of physical background,
although it reduces the detector leakage current.

The detector was read out through two 5 m long RG62 cables.
Each cable was additionally shielded with an aluminum wire braid.

\section{Measurement at FNG}\label{sec:fng}
The detector response to DT neutrons was measured at Frascati Neutron Generator (FNG) of ENEA~\cite{fng_facility}.
The FNG is a neutron source based on the T(d,n)$\alpha$ fusion reaction which generates up to
$5\times 10^{10}$ 14-15 MeV n/s.
The source strength is monitored measuring the produced $\alpha$ particles associated with the DT fusion reaction. 
The $\alpha$ particles are detected in a small silicon surface barrier detector installed inside the beamline.
At the nominal beam current the resulting uncertainty on the source strength measurement is lower than 4\%.
The TiT target is located at about 4 m above the floor and at more than 4 m from hall's walls and ceiling,
in order to reduce the neutron background from backscattering.

\subsection{Experiment}
\begin{figure}[!ht]
\begin{center}
\includegraphics[bb=2.5cm 1.5cm 20cm 27.5cm, scale=0.35, angle=270]{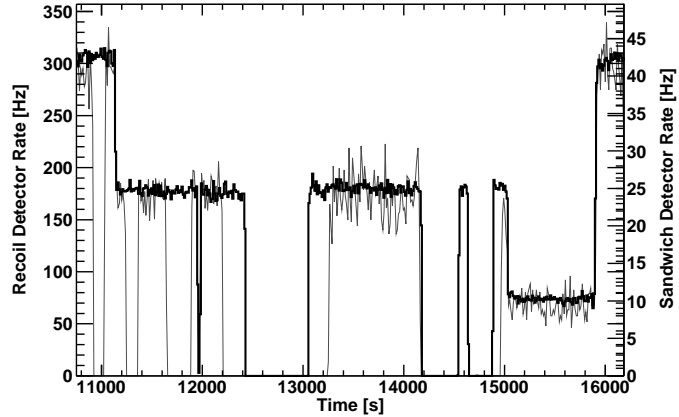}
\caption{\label{fig:fng_a_recoil_swd_rates}Rate of the FNG $\alpha$ recoil monitor (thick black histogram)
in comparison to the sandwich detector rate (thin grey histogram) rescaled for comparison.
The periods of zero rate in the sandwich detector correspond to the pauses between runs.}
\end{center}
\end{figure}

The sandwich detector was installed at 6 cm from the TiT target at 90 degrees with respect to the impinging $D$ beam.
During this experiment FNG was operated in DT-mode with a beam current of 0.2 mA and a total
14-15 MeV neutron yield of $2.2\times 10^{10}$ n/s.
The number of $\alpha$ recoil monitor counts were recorded each 3 s to extract the total neutron yield.
The corresponding source strength profile during the experiment is shown in Fig.~\ref{fig:fng_a_recoil_swd_rates}.
The sandwich detector rate, measured in six separate runs, followed the recoil monitor rate very closely.

\begin{figure}[!ht]
\begin{center}
\includegraphics[bb=2.5cm 1.5cm 20cm 27.5cm, scale=0.35, angle=270]{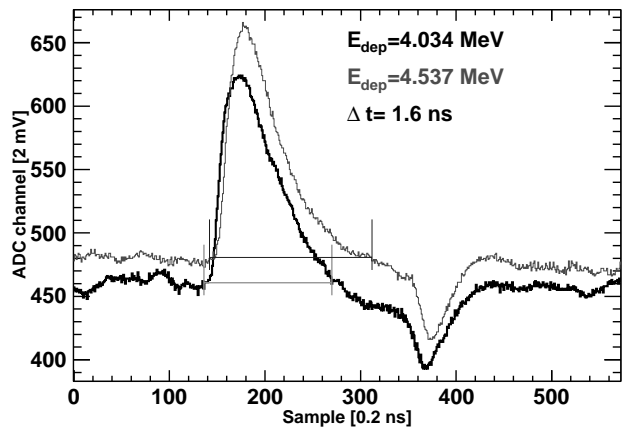}
\caption{\label{fig:wf_coin_event}Waveforms of two readout channels of the sandwich detector
for a typical coincidence event near $^{12}C(n,\alpha)^{9}Be$ peak. The straight lines
indicate the corresponding intervals of integration used to obtain the deposited energy values.
Obtained deposited energies $E_{dep}$ and delay between two signals $\Delta t$ are given.}
\end{center}
\end{figure}

The two sandwich detector outputs were connected via the 5 m RG62 cables to custom fast charge amplifiers~\cite{cardarelli_amp}. At the amplifier inputs fast DC decouplers were
used to connect the detector bias voltage of -80 V,
corresponding to 1.6 V/$\mu$m, supplied by an Ortec 710 module~\cite{ortec}.
The detector signals boosted and shaped in the primary amplifiers
were further amplified by a Phillips Scientific 771 amplifier~\cite{ps} with voltage gain set to $4$.
The overall gain of this chain was about 100 pVs/fC.
The amplified signals were sent to a SIS3305 10-bit digitizer~\cite{sis} working in 5 Gs/s mode.
The digitizer internal trigger was configured to fire on a coincidence~\cite{amp_preprint}
between the two detector channels within a 64 ns interval.
The individual channel threshold was set to 40 mV.
An example of waveforms measured in a single event is shown in Fig.~\ref{fig:wf_coin_event}.
The Data AcQuisition (DAQ) system was assembled using VME modular electronics.
It was based on 
a Concurrent Tech. VX813-09x single board computer~\cite{concur}, used as the VME controller and acquisition host.
The SIS3305 was configured to place interrupts on the VME bus when the number of events in its 2 GB buffer reached 16.
Then the data were copied from the SIS3305 buffer by means of a fast DMA MBLT transfer and copied to a secondary DAQ thread.
This secondary DAQ thread saved the data on a fast Compact Flash card.
For every event $N_{samples}=960$ samples were saved for each of two diamonds, corresponding to a total
waveform duration of 192 ns. Within these samples the actual signal length was $L_{signal}\sim 140-220$ samples or 30-45 ns.
Two TDC values of the common trigger time as seen in the two channels were also saved.
For each event the data size amounted to 240 kB acquired at a rate below 45 Hz,
therefore with negligible DAQ dead time.

\subsection{Data analysis}
For the absolute normalization of the data the FNG $\alpha$ recoil monitor system combined with the
rate-to-flux conversion factor obtained from MCNP calculations were used.
The MCNP model of FNG facility describes in detail its geometry and was accurately
tested in Ref.~\cite{fng_mcnp}.

\begin{figure}[!ht]
\begin{center}
\includegraphics[bb=2.5cm 1.5cm 20cm 27.5cm, scale=0.35, angle=270]{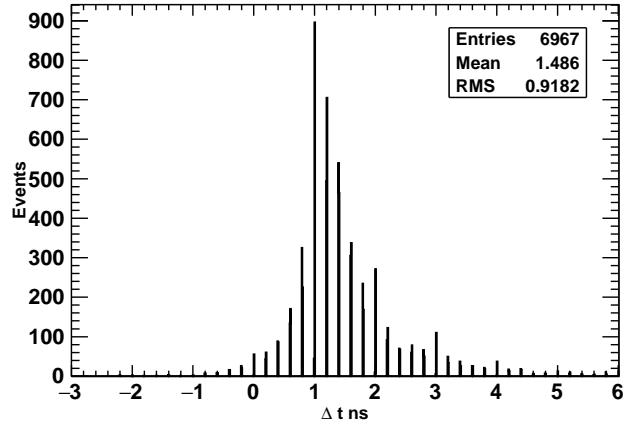}
\caption{\label{fig:cdf_dt_run39}Difference between arrival times of the signals from the two detector channels.}
\end{center}
\end{figure}

Significant EMI bursts were observed during the experiment.
Although the 64 ns narrow coincidence trigger was fairly efficient in suppressing this noise,
the saved data were further filtered to discard remaining backgrounds.
This procedure included the following checks:
the integral of the first 100 samples before the signal was checked to be equal to the baseline within 3$\sigma$,
the trigger sample was checked to lie at the sample number $>50$ and $<N_{samples}-L_{signal}-50$,
the difference of arrival times of the signals from the two channels was requested to lie within 6 ns
(see Fig.~\ref{fig:cdf_dt_run39}),
and finally the integral of the signal had to be above a software threshold (0.5 MeV) in both channels.
A higher threshold (e.g. 4 MeV to suppress signals from the elastic scattering on carbon)
would allow for a better signal-to-background ratio,
but in the present study we were also interested to establish the amount of background.

%
%

In order to calibrate the ADC energy scale a polyethylene moderator was installed behind the detector in the last two runs.
The calibration of the energy deposited in each crystal was performed as in Ref.~\cite{sdw_calib}.
This procedure exploited the highly exothermic reaction of thermal neutrons $^{6}Li(n,t)\alpha$
in 100 nm thick LiF layer interposed between the two diamonds.
The energy of the produced $t$, corrected for the energy lost in LiF layer and metallic anode,
was used as a calibration reference. In order to have a second calibration reference,
the digitizer baseline was taken as the zero energy point.
The RMS of the $t$-peak $\sigma_{t from ^{6}Li}$ was found to be $280$ keV. This value is three times larger than the one
measured in Ref.~\cite{sdw_tapiro} at the TAPIRO reactor (Casaccia, ENEA) with the same amplifiers and seven times worse than
the resolution obtained with standard charge sensitive amplifiers (Silena 205)~\cite{sdw_calib} at the TRIGA reactor (Pavia, LENA).
Such a difference confirms that the dominant contribution to the resolution
was due to the high frequency EM noise in FNG hall.

%
\begin{figure}[!ht]
\begin{center}
\includegraphics[bb=2.5cm 1.5cm 20cm 27.5cm, scale=0.35, angle=270]{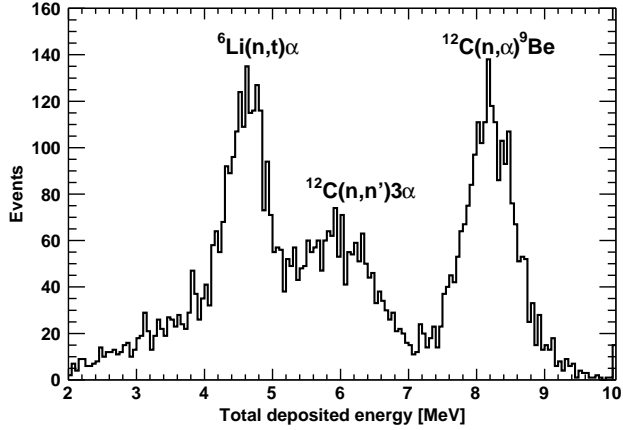}
\caption{\label{fig:edep_tot_run39}Total deposited energy in the two crystals of the sandwich detector:
left peak is due to $^{6}Li(n,t)\alpha$ reaction of thermal neutrons in the interposed LiF layer,
right peak is due to $^{12}C(n,\alpha)^{9}Be$ reaction of 14 MeV neutrons,
the background in the middle is produced via $^{12}C(n,n')3\alpha$ reaction by 14 MeV neutrons.}
\end{center}
\end{figure}

The energies deposited in the two diamonds were summed up to measure
the total deposited energy showed in Fig.~\ref{fig:edep_tot_run39}.
The obtained distribution exhibits three main structures indicating different neutron interaction mechanisms.
A large peak at 4.7 MeV is produced by thermal neutron interaction with interposed LiF layer $^{6}Li(n,t)\alpha$ ($Q=4.7$ MeV).
This reaction was used for energy calibrations and it was observed only during calibration runs
with moderator polyethylene cylinder located near the detector.
The peak RMS $\sigma_{^{6}Li(n,t)\alpha}$ was found to be $370$ keV in agreement with the quadratic combination
of the two diamond resolutions.
The other two structures are due to 14 MeV neutron interaction in the diamond bulk.
The separated peak at 8.3 MeV is produced via the $^{12}C(n,\alpha)^{9}Be$ reaction,
where the whole neutron kinetic energy is converted into the energy of charged particles and measured.
The RMS of the 8.3 MeV peak $\sigma_{^{12}C(n,\alpha)^{9}Be}$ was $390$ keV, slightly larger than for the 4.7 MeV peak.
The difference between these values is related to the different location of the interaction point
within the detector and therefore the different amount of material crossed by $\alpha$ particles.
In average the $\alpha$ particles produced through $^{12}C(n,\alpha)^{9}Be$ reaction in the diamond bulk
had to cross twice more material than the ones from the $^{6}Li(n,t)\alpha$ reaction generated in the interposed LiF layer.
Combining these RMS values it is possible to estimate the intrinsic resolution
of the detector due to stochastic energy loss of $\alpha$ particles in the inactive materials
$\sigma_{\alpha from ^{12}C}$:
\begin{equation}
\sigma^2_{^{12}C(n,\alpha)^{9}Be}-\sigma^2_{^{6}Li(n,t)\alpha} \sim 
\sigma^2_{\alpha from ^{12}C} - \Bigl(\sigma^2_{\alpha from ^{6}Li} - \sigma^2_{t from ^{6}Li}\Bigr) ~.
\end{equation}
\noindent The numerical evaluation gives $\sigma_{\alpha from ^{12}C}\sim 70$ keV
in agreement with Geant4 simulations. Adding quadratically to this value the minimal
resolution of the readout system from Ref.~\cite{sdw_calib} (73 keV with similar detector)
one expects to achieve RMS of $^{12}C(n,\alpha)^{9}Be$ peak of 100 keV
if the EM noise was eliminated.

\subsection{Backgrounds}\label{sec:bkg}
The background of this measurement is mostly composed of EM noise
and some accidental coincidences of elastic n-C scattering. Both these
backgrounds were removed offline by requiring a 6 ns maximum time difference
between the two diamonds as shown in Fig.~\ref{fig:cdf_dt_run39}.

There was also a part of the neutron spectrum outside the main 14 MeV peak
due to neutrons scattered from surrounding materials. This component of the spectrum was included
in the Monte Carlo simulation.

\subsection{Systematic uncertainties}\label{sec:sys_err}
The present measurement served as an exploratory test of the detector response
to 14 MeV neutrons.
Thus, we considered only one major systematic uncertainty of the data.
This comes from the determination of the neutron flux
impinging on the face of the detector.
The overall neutron yield of FNG obtained by means of the $\alpha$ recoil monitor
counts had an uncertainty of 4\%. However, the detector position with respect to the TiT
target was determined with an average precision of 1 cm. Furthermore, the MCNP model of FNG target
for the used detector position cannot describe the neutron flux map to better than 10\%.
The combination of these uncertainties was estimated to be 30\%.

\section{Results}\label{sec:res}
\begin{figure}[!h]
\begin{center}
\includegraphics[bb=2.5cm 0cm 20cm 26cm, scale=0.35, angle=270]{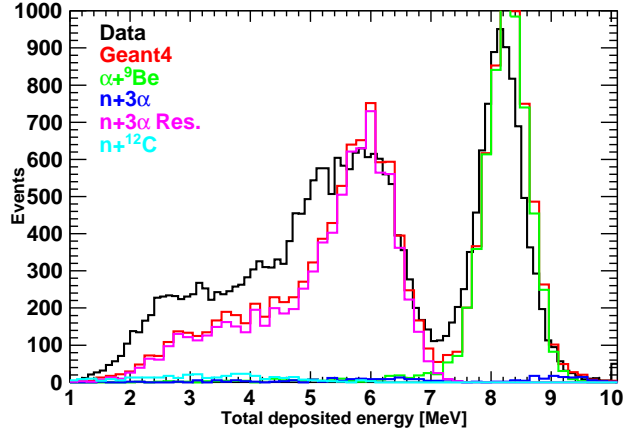}
\caption{\label{fig:fng_swd2_etot}Total energy deposited in the sandwich detector (black histogram)
in comparison with Geant4 simulations (red histogram).
For Geant4 simulations various reaction contributions are also shown:
green - $^{12}C(n,\alpha)^9Be$ reaction,
blue - $^{12}C(n,n')3\alpha$ continuum reaction,
magenta - $^{12}C(n,n')3\alpha$ resonance reaction,
cyan - $^{12}C(n,n')$ reaction.}
\end{center}
\end{figure}

\begin{figure}[!h]
\begin{center}
\includegraphics[bb=3.5cm 0cm 20cm 26cm, scale=0.35, angle=270]{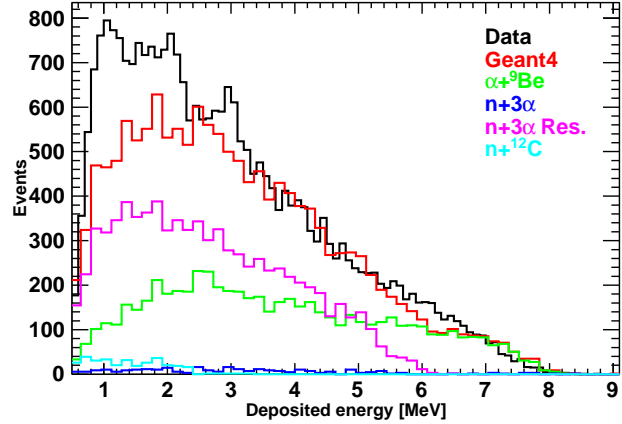}
\caption{\label{fig:fng_swd2_e2}Energy deposited in the first diamond of sandwich detector
for coincidence events
in comparison with Geant4 simulations (red histogram).
Colors are the same as in Fig.~\ref{fig:fng_swd2_etot}.}
\end{center}
\end{figure}

The measured total and single diamond deposited energy distributions were compared with
Geant4.9.5.p01~\cite{geant4} simulations performed with a realistic neutron spectrum (see Ref.~\cite{fng_p_recoil}).
In these comparisons, shown in Figs.~\ref{fig:fng_swd2_etot} and~\ref{fig:fng_swd2_e2},
the Geant4 events were normalized to the integrated neutron flux $\int \phi_n^{tot} dt$
at the detector face obtained by means of FNG recoil monitor.
To this end all simulated Geant4 spectra were rescaled by the factor:
\begin{equation}
L_{MC}= \int \phi_n^{tot} dt \frac{S_{source}}{N_{gen}} ~,
\end{equation}
\noindent where $N_{gen}$ is the total number of neutrons generated on the surface of area $S_{source}$.
Geant4 events were also separated into four major contributions
due to different reactions induced by 14 MeV neutron in the detector.

\begin{figure}[!h]
\begin{center}
\includegraphics[bb=2.5cm 0cm 20cm 26cm, scale=0.35, angle=270]{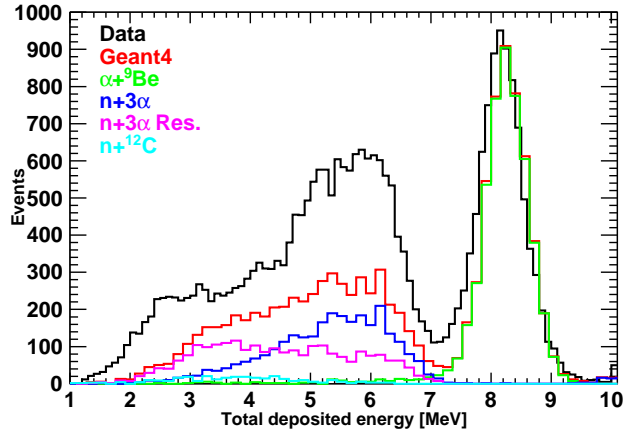}
\caption{\label{fig:fng_swd2_etot_n3a}The same as in Fig.\ref{fig:fng_swd2_etot},
but direct $^{12}C(n,n')3\alpha$ continuum reaction is used in Geant4 model $MT=91$.}
\end{center}
\end{figure}

The relevant feature of the total deposited energy spectrum is the peak
at 8.3 MeV produced by complete neutron energy conversion through $^{12}C(n,\alpha)^9Be$ reaction.
The peak is separated by 2 MeV from all other reactions thanks to the lower $Q$-value.
This is the reaction often used to determine ion temperature in fusion reactor
and to monitor fusion rate. The peak width gives DT neutron energy resolution (intrinsic energy spread at FNG is 165 keV),
equal to 870 keV (FWHM) in the present measurement. But only 30\% of the width
was intrinsic to the detector (see also Ref.~\cite{sdw_calib}), while the rest was related to high EM noise.

The rest of the spectrum is dominated by a physical background due to the reaction $^{12}C(n,n')3\alpha$.
In this reaction a part of energy is taken by the undetected final state neutron $n'$,
making impossible initial energy reconstruction.
According to Refs.~\cite{n3a_frey,n3a_antolkovic} the reaction $^{12}C(n,n')3\alpha$
proceeds mostly through formation of various excited states
$^{12}C(n,n')^{12}C^*(7.65, 9.64, ... MeV)$ and $^{12}C(n,\alpha)^9Be^*(2.43 MeV)$,
which decay $^{12}C^*\to 3\alpha$ and $^9Be^*\to n+2\alpha$, respectively.
The direct, non-resonant break-up reaction $^{12}C(n,n')3\alpha$ contribution is small.
ENDF data used in the NeutronHP physics model of Geant4 include the explicit
contributions of $^{12}C^*(7.65, 9.64, ... MeV)$ excited states.
Instead, all the rest of the cross section is attributed to the continuum inelastic channel $MT=91$.
Attributing all this continuum part of $^{12}C(n,n')3\alpha$ reaction
to $^{12}C(n,\alpha)^9Be^*(2.43 MeV)$ channel allows to describe the data fairly well.
On the contrary, generating the continuum channel $MT=91$ through phase space distribution
of direct four-body break-up as shown in Fig.~\ref{fig:fng_ddl_edep_n3a}
underestimates the data in the region of 6 MeV by a factor of two.
Even rescaling this contribution by a factor of two does not allow to describe our spectra,
in particular in the region 3.5-4.5 MeV the simulation will overestimate the data.
Not to mention that the factor of two increase of continuum part of $^{12}C(n,n')3\alpha$ cross section
would be in disagreement with Refs.~\cite{n3a_frey,n3a_antolkovic}.

\begin{figure}[!ht]
\begin{center}
\includegraphics[bb=3.5cm 0cm 20cm 26cm, scale=0.35, angle=270]{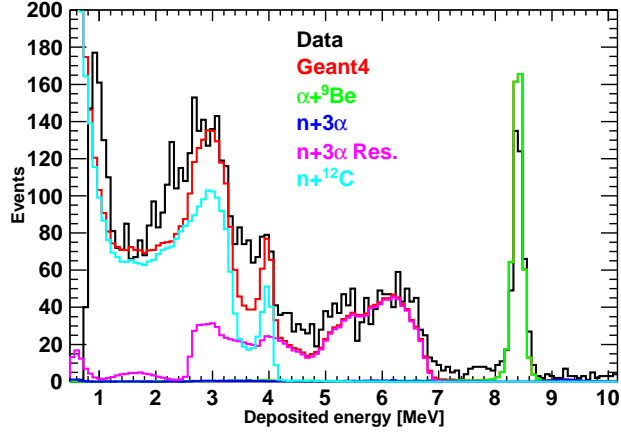}
\caption{\label{fig:fng_ddl_edep}Energy deposited in a single, independent diamond detector (black histogram)
from Ref.~\cite{fng_p_recoil} in comparison with Geant4 simulations (red histogram).
For Geant4 simulations various reaction contributions are also shown:
green - $^{12}C(n,\alpha)^9Be$ reaction,
blue - $^{12}C(n,n')3\alpha$ continuum reaction,
magenta - $^{12}C(n,n')3\alpha$ resonance reaction,
cyan - $^{12}C(n,n')$ reaction.}
\end{center}
\end{figure}

\begin{figure}[!ht]
\begin{center}
\includegraphics[bb=3.5cm 0cm 20cm 26cm, scale=0.35, angle=270]{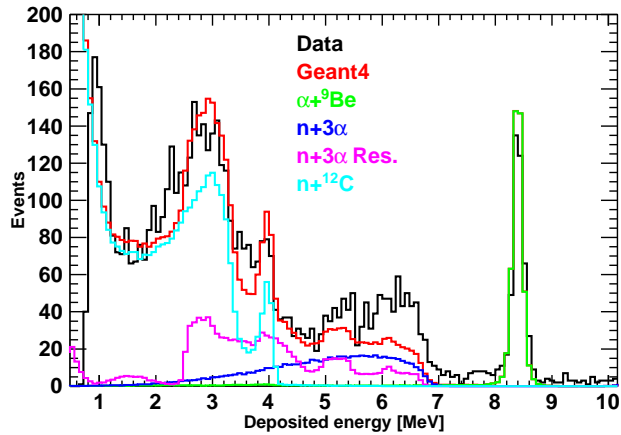}
\caption{\label{fig:fng_ddl_edep_n3a}The same as in Fig.\ref{fig:fng_ddl_edep},
but direct $^{12}C(n,n')3\alpha$ continuum reaction is used in Geant4 model $MT=91$.}
\end{center}
\end{figure}

A similar agreement can be seen in the spectrum measured in Ref.~\cite{fng_p_recoil} on a single, independent
diamond sensor without requiring coincidences shown in Figs.~\ref{fig:fng_ddl_edep} and \ref{fig:fng_ddl_edep_n3a}.
Indeed using direct break-up phase space distribution underestimates the data
in the same region of 3.5-4.5 MeV. Instead, ascribing the $MT=91$ strength to
$^{12}C(n,\alpha)^9Be^*(2.43 MeV)$ channel allows to describe the data fairly well.


The same reaction, going through the production of $^{12}C$ excited states decaying in $3\alpha$,
gives a much smaller contribution. It exhibits several steps in the spectra
due to decay of the higher $^{12}C$ levels at rest. For example, the most evident step at 2.4 MeV
in Fig.~\ref{fig:fng_ddl_edep} is due to the decay of 9.64 MeV level.

The contribution of two consecutive elastic scatterings from $^{12}C$ into
ground or first excited state of $^{12}C$ gives a very small contribution.
This physical background can be further reduced by increasing the thresholds
and by reducing the diamond thickness. The contribution of accidental coincidences
of elastic scatterings off $^{12}C$ was neglected because for a 6 ns coincidence
interval its contribution is $<10^{-5}$.

\begin{figure}[!h]
\begin{center}
\includegraphics[bb=3.5cm 0cm 20cm 26cm, scale=0.35, angle=270]{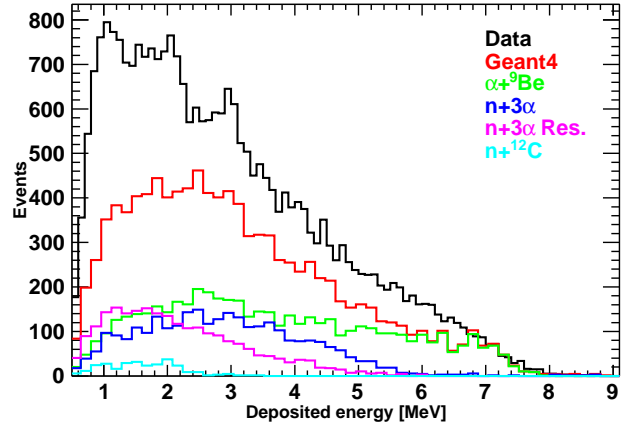}
\caption{\label{fig:fng_swd2_e2_n3a}The same as in Fig.\ref{fig:fng_swd2_e2},
but direct $^{12}C(n,n')3\alpha$ continuum reaction is used in Geant4 model for $MT=91$.}
\end{center}
\end{figure}

The energy spectrum in a single diamond measured in coincidence, shown in Figs.~\ref{fig:fng_swd2_e2}
and \ref{fig:fng_swd2_e2_n3a}, suggests similar conclusions.
Moreover, from all three distributions it is clear that the simulations underestimate the data
in the region around 1-2.5 MeV for the single diamond spectra and 2-5 MeV for the total deposited energy.
This deficit of strength around 5 MeV is probably related to the contribution of $^9Be^*(3.05 MeV)$ excited state,
while in the region 2-4 MeV is due to non-uniformity of Cr and LiF layers
allowing lower energy $\alpha$ particles to trigger the coincidences.
From the Fig.~\ref{fig:fng_swd2_e2} it is also following that the physical background events
deposit less energy in the single crystal, in particular the resonance part of the $^{12}C(n,n')3\alpha$ reaction
and the elastic n-C scattering. Therefore, increasing the single crystal threshold to about 2.5 MeV could
allow to reduce the background by a factor of eight with 50 \% efficiency loss.

These data allow to determine 14 MeV neutron detection efficiency through $^{12}C(n,\alpha)^9Be$ reaction.
Taking the number of events in the peak and dividing it by the total accumulated neutron flux
on the detector face we obtain a sensitivity of $5\times 10^{-7}$ counts cm$^2$/n
for $3\times 3$ mm$^2$ detector.
In order to compare this with analytic estimate we evaluated in Geant4 simulations
the diamond active volume. This volume is defined by the maximum depth of the first diamond
from which the produced $\alpha$ particle could emerge and hit the second crystal.
Simulations indicate a maximum depth of 22 $\mu$m for 1 MeV threshold.
Combining this value with the known $^{12}C(n,\alpha)^9Be$ cross section (70 mb at 14 MeV)
and assuming 25 \% probability for the $\alpha$ to enter into the second crystal and to trigger the coincidence
we find good agreement with the measured rate of 13 Hz vs. 15 Hz estimated
at $2.7\times 10^7$ n/cm$^2$s.

\section{Conclusions}\label{sec:conclusions}
We measured the response of the sandwich diamond detector developed in Ref.~\cite{sdw_calib}
to 14 MeV neutrons from a DT generator.
The neutrons were measured through $^{12}C(n,\alpha)^{9}Be$ reaction in the first diamond's bulk,
where the produced $\alpha$ traveled into the second diamond, triggering the coincidences.
The choice of $50$ $\mu$m thin diamond crystals is expected to increase the detector radiation
hardness up to $>10^{16}$ n/cm$^2$ satisfying ITER RNC requirements.
The measured detector response was found to be in good agreement with Geant4 simulations.
The neutron energy resolution deduced from $^{12}C(n,\alpha)^{9}Be$ peak width was found
to be 870 keV (FWHM), but, taking into account previous measurements performed with the same detector,
only 240 keV were intrinsic to the detector design.
In particular, $\alpha$ energy loss fluctuations in the detector's metallic contacts and
intermediate LiF layer represented its intrinsic resolution.
This intrinsic resolution adds only 10\% to ITER DT neutron energy spread of 500 keV.
The remaining resolution was due to the environmental EM noise
induced in the 5 m cable between detector and the first amplifier.

The relatively small detector active volume leads to a lower detector efficiency
with respect to typical 500 $\mu$m thick crystals.
This renders the detector suitable for neutron fluxes $>10^7$ n/cm$^2$s.

The comparison of the data to Geant4 simulations allowed to ascribe
the continuum inelastic contribution (ENDF identification $MT=91$)
of $^{12}C(n,n')3\alpha$ reaction to $^{12}C(n,\alpha)^9Be^*(2.43 MeV)$ channel
in agreement with Refs.~\cite{n3a_frey,n3a_antolkovic}.
The direct four-body break-up through phase space distribution of reaction products
cannot accommodate our data.

\section*{Acknowledgements}
The authors would like to acknowledge the support provided during the experiments
by the staff and technical services of FNG facility.
This work was supported by the Istituto Nazionale di Fisica Nucleare INFN-E project.

\bibliographystyle{elsarticle-num}
\bibliography{sdw_dt}

\end{document}